\documentclass[letter]{aa} 

\usepackage{graphicx}
\usepackage{txfonts}
\begin{document}

   \title{Spectro-astrometry of the pre-transitional star LkCa 15 does not reveal an accreting planet but extended H$\alpha$ emission}
   
   \author{I. Mendigut\'\i{}a\inst{1}
   \and
   R.D. Oudmaijer\inst{2}
   \and
   P.C. Schneider\inst{3}
   \and
   N. Hu\'elamo\inst{1}
   \and
   D. Baines\inst{4}
   \and
   S.D. Brittain\inst{5}
   \and
   M. Aberasturi\inst{6}
   }          

   \institute{$^{1}$Centro de Astrobiolog\'{\i}a (CSIC-INTA), Departamento de Astrof\'{\i}sica, ESA-ESAC Campus, PO Box 78, 28691 Villanueva de la Ca\~nada, Madrid, Spain. \email{imendigutia@cab.inta-csic.es}\\
   $^{2}$School of Physics and Astronomy, University of Leeds, Woodhouse Lane, Leeds LS2 9JT, UK.\\
   $^{3}$Hamburger Sternwarte, Gojenbergsweg 112, 21029 Hamburg, Germany.\\
   $^{4}$Quasar Science Resources, European Space Astronomy Centre, PO Box 78, 28691 Villanueva de la Ca\~nada, Madrid, Spain.\\
   $^{5}$Department of Physics \& Astronomy, 118 Kinard Laboratory, Clemson University, Clemson, SC 29634-0978, USA.\\
   $^{6}$Serco, European Space Astronomy Centre, PO Box 78, 28691 Villanueva de la Ca\~nada, Madrid, Spain.\\}

   \date{Received September 13, 2018; accepted October 4, 2018}

 
  \abstract
   {The detection of forming planets in protoplanetary disks around young stars remains elusive, and state-of-the-art observational techniques provide somewhat ambiguous results. The pre-transitional T Tauri star LkCa 15 is an excellent example. It has been reported that it could host three planets;  candidate planet b is in the process of formation, as inferred from its H$\alpha$ emission. However, a more recent work casts doubts on the planetary nature of the previous detections.}
   {We  test the potential of spectro-astrometry in H$\alpha$ as an alternative observational technique to detect forming planets around young stars, taking LkCa 15 as a reference case}
   {LkCa 15 was observed with the ISIS spectrograph at the 4.2m William Herschel Telescope (WHT). The slit was oriented towards the last reported position of LkCa 15 b (parallel direction) and 90$\degr$ from that (perpendicular). The photocenter and full width half maximum (FWHM) of the Gaussians fitting the spatial distribution at H$\alpha$ and the adjacent continuum were measured. A well-known binary (GU CMa) was used as a calibrator to test the spectro-astrometric performance of ISIS/WHT.}
   {A consistent spectro-astrometric signature is recovered for GU CMa. However, the photocenter shift predicted for LkCa 15 b is not detected, but the FWHM in H$\alpha$ is broader than in the continuum for both slit positions. Our simulations show that the photocenter and FWHM observations cannot be explained simultaneously by an accreting planet, but the lack of photocenter shift alone could still be consistent with an emitting planet with contrast $\gtrsim$ 5.5 mags in H$\alpha$ or $\lesssim$ 6 mag in the adjacent continuum. In turn, both spectro-astrometric observations are naturally reproduced from a roughly symmetric H$\alpha$ emitting region centered on the star and extent comparable to the orbit originally attributed to the planet at several au.} 
   {The extended H$\alpha$ emission around LkCa 15 could be related to a variable disk wind, but additional multi-epoch data and detailed modeling  are necessary to understand its physical nature. Optical spectro-astrometry carried out with mid-size telescopes is capable of probing small-scale structures in relatively faint young stars that are not easily accessible with state-of-the-art instrumentation mounted on larger telescopes. Therefore, spectro-astrometry in H$\alpha$ is able to test the presence of accreting planets and can be used as a complementary technique to survey planet formation in circumstellar disks.}

\titlerunning{Spectro-astrometry of the pre-transitional star LkCa 15}
   \maketitle
%

\section{Introduction}
\label{Sect:Intro}
The detection of close-orbit ($<$ 100 au) forming planets in protoplanetary disks around young stars is a major goal in current astrophysical research. An increasing number of such detections will allow us to carry out comparative studies with the wide number of exoplanets detected around mature stars, significantly improving the empirical understanding on the formation and evolution of planetary systems. However, the current situation is still far from this scenario. Relatively classical methods applied to young stars are providing some detections of possible substellar/planetary companions from photometric transits or radial velocity Doppler signatures  \citep[e.g.,][]{Donati16,Krull16,Almeida17,Yu17,Osborn17}. Nonetheless, the stellar activity is a limitation for these techniques, which can be applied mainly to detect very close massive companions or when the stellar host is a comparatively evolved weak-line T Tauri star, for which the main phase of planet formation has most probably finished.  High-resolution imaging, differential polarimetry, and interferometric/sparse aperture masking techniques applied with state-of-the-art instrumentation are also providing promising results for a few young planet candidates \citep[e.g., the recent results in][]{Pinte18,Teague18,Keppler18,Wagner18}. However, the complexity of the data reduction and the interpretation involved has partially resulted in a discussion on the real nature of the detections \citep[see, e.g., the cases of \object{HD 100546} and \object{MWC 758} in][and references therein]{Follette17,Rameau17,Mendi17b,Huelamo18}. 

A representative example of this debate is the T Tauri star \object{LkCa 15} \citep[M$_*$ $\sim$ 1.2M$_{\odot}$, R$_*$ $\sim$ 1.5R$_{\odot}$, T$_*$ $\sim$ 4900K;][]{Manara14}, located  159 $\pm$ 1 pc from us according to the recent Gaia Data Release \citep{Lindegren18}. LkCa 15 is a pre-transitional star with a dusty inner disk extending up to $\sim$ 30 au from the star, a possibly misaligned outer disk extending from $\sim$ 60 au, and a gap void of dust in between \citep[see, e.g.,][and references therein]{Pietu07,Thalmann15,Thalmann16,Oh16}. First, \citet{Sallum15} built on the initial results by \citet{Kraus12} and reported the detection of three planets at orbital radii of $\sim$ 15 au based on near-IR non-redundant masking with the Large Binocular Telescope. In particular, candidate planet b appears to show additional H$\alpha$ emission consistent with accretion, according to simultaneous differential imaging carried out by the same authors with the Magellan Adaptive Optics System. In contrast, \citet{Thalmann16} analyzed scattered light with SPHERE at the Very Large Telescope (VLT) and conclude that the IR-bright candidate planets could be persistent structures in the inner dust disk. However, \citet{Thalmann16} left the door open to the possible presence of LkCa 15 b due to the independent detection in H$\alpha$. In summary, LkCa 15 is an excellent example of the fact that complementary observational techniques are necessary to test the presence of forming planets around young stars and eventually provide new detections from alternative approaches.         

Spectro-astrometry is a robust observational technique able to probe physical structures including disks, outflows, and binaries with a brightness contrast of several magnitudes at angular scales of (sub-)mas \citep[see, e.g., the reviews in][]{Bailey98a,Whelan08,Brittain15}. This method exploits the spatial information in a 2D spectrum by measuring the photocenter and full width half maximum \citep[FWHM; for details about this parameter in spectro-astrometry see, e.g.,][]{Baines06,Wheelwright10} of the Gaussians characterizing the observed spatial distribution at different wavelengths, providing information on the relative brightness, position, extent, and possible asymmetries of the different structures within the source under analysis. In fact, optical spectro-astrometry has demonstrated enormous capabilities of detecting stellar binaries in T Tauri and Herbig Ae/Be stars when the H$\alpha$ contrast between the primary and the secondary is different than the contrast in the continuum \citep[e.g.,][]{Bailey98b,Takami03,Baines06,Wheelwright10}. A particularly relevant advantage of this technique is that it allows the recovery of the individual intensity spectra of each component in a double system when photocenter shifts are observed \citep[e.g.,][and references therein]{Bailey98a,Wheelwright10}, which would have obvious applications in  understanding  the accreting phase of forming planets if their H$\alpha$ emission profile could be extracted. Based on the first detection by \citet{Kraus12}, \citet{Whelan15} applied X-Shooter/VLT spectro-astrometry aiming to test for the presence of LkCa 15 b. However, because of X-Shooter artifacts and instrumental distortions of the point spread function (PSF), the non-detection of a spectro-astrometric signal was not sufficient to rule out the presence of the companion. Moreover, the information provided by \citet{Sallum15} on the H$\alpha$ emission of LkCa 15 b was not available at that time. From the H$\alpha$ contrast and separation between LkCa 15 and LkCa 15 b \citep[c$_{H\alpha}$ $\sim$ 5.2 magnitudes, $s$ $\sim$ 93 mas;][]{Sallum15} the expected photocenter shift is $\delta$phot = L$_p$$\times$$s$/(L$_*$ + L$_p$) = $s$/(10$^{0.4c_{H\alpha}}$+1) $\sim$ 0.8 mas. This is well below the accuracy reached by \citet{Whelan15} (see, e.g., the bottom panels of their Fig. 3), thus the detection of LkCa 15 b was impossible from that data.

Our aim is to test the potential of spectro-astrometry to detect forming planets, taking LkCa 15 as a reference case. Section \ref{Sect:data_reduction} shows the data reduction and results, which are analyzed in Sect. \ref{Sect:analysis}. The summary and conclusions are presented in Sect. \ref{Sect:conclusions}.

\section{Data reduction and results}
\label{Sect:data_reduction}
Observations were carried out in visitor mode with the ISIS spectrograph at the 4.2m William Herschel Telescope (WHT) in La Palma Observatory during the first half night of 4 January 2018. The R1200R grating centered at H$\alpha$ ($\Delta$$\lambda$ $\sim$ 700 $\AA{}$; Fig. \ref{Fig:fullspec}) was used with a nominal resolution of 0.26 $\AA$ pixel$^{-1}$ and a plate scale of 0.22$\arcsec$ pixel$^{-1}$. The spectral resolution inferred from the FWHM of the lines in the calibration lamps is $R$ = $\lambda$/$\delta$$\lambda$ $\sim$ 9500, or $\sim$ 30 km s$^{-1}$ at H$\alpha$. The average seeing ($\pm$3$\sigma$) was $\sim$ 1.2$\pm$ 0.5$\arcsec$, as estimated from the PSF given by the continuum adjacent to the H$\alpha$ emission spectra of LkCa 15 (see also Table \ref{table:spectra}).

\subsection{GU CMa as a test case}
\label{Sect:gucma}
Apart from LkCa 15 (R = 11.6 mag), the young and bright binary star \object{GU CMa} (R = 7.5 mag) was observed during the same night, with the same instrumental configuration and observing strategy. These and the corresponding data reduction are described in the following since GU CMa was used as a calibrator to test the spectro-astrometric performance of ISIS/WHT. 

\begin{figure}
\centering
 \includegraphics[width=9cm,clip=true]{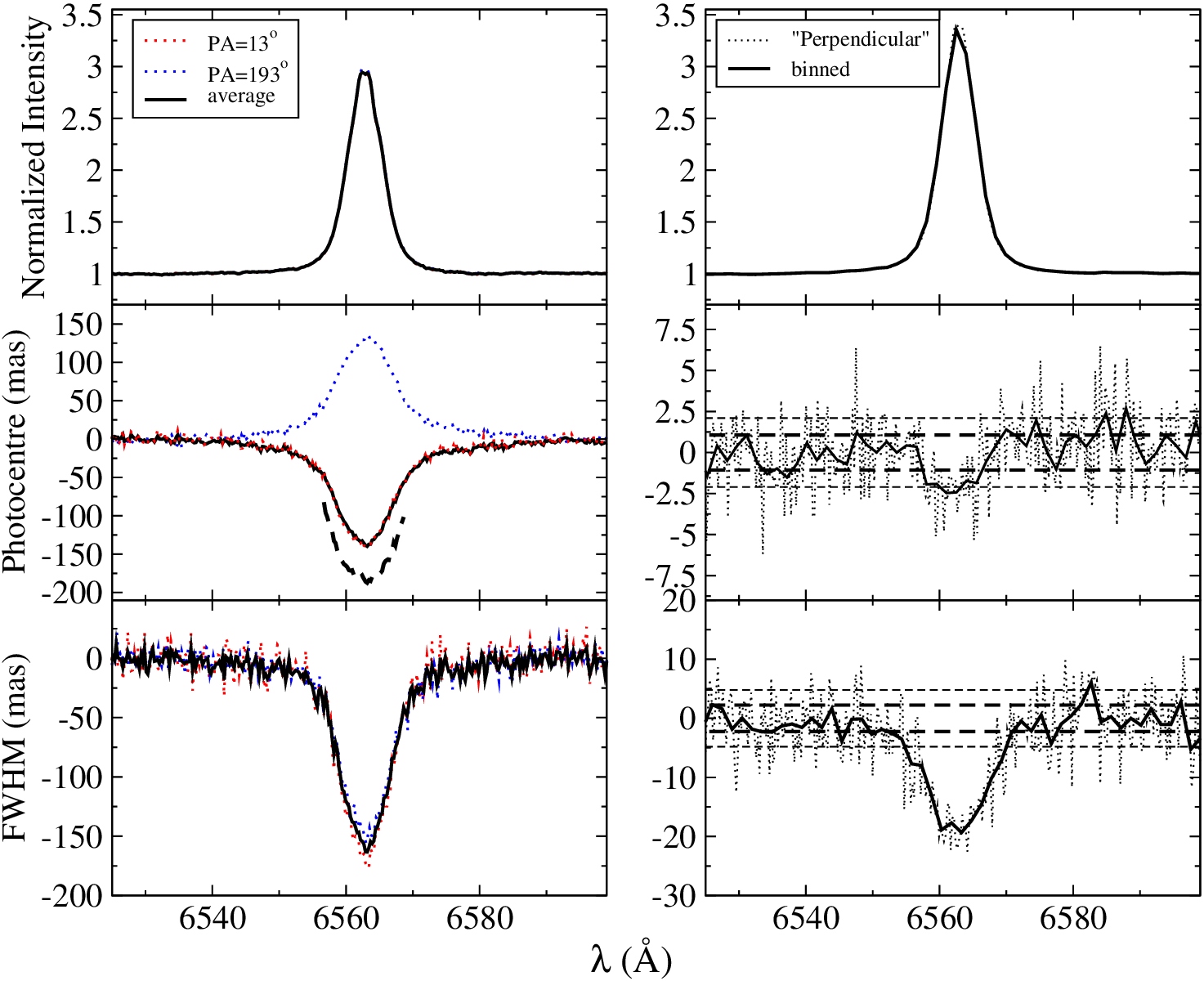}
\caption{Intensity, photocenter, and FWHM spectra of GU CMa from top to bottom for the parallel (left) and perpendicular (right) slit directions. The left panels plot the individual spectra for the two parallel positions and the averaged spectra (see legend). The continuum-corrected photocenter shift is indicated with the black dashed line in the middle left panel. The right panels show the averaged spectra for the perpendicular orientation ($PA$ = 103$\degr$ and 283$\degr$) with the original spectral resolution (dotted lines) and after rebinning (solid lines). The corresponding $\pm$ 1$\sigma$ levels are shown in the middle and bottom right panels with the horizontal normal and bold dashed lines, indicating $\pm$ 2.1, $\pm$ 1.1, and $\pm$ 4.8, $\pm$ 2.2 mas for the photocenter and FWHM accuracies.}
\label{Fig:gu_cma}
\end{figure}

Centered on the primary star, the slit (width = 1$\arcsec$, length = 4$\arcmin$) was oriented in the parallel direction, i.e., first aligned towards the secondary \citep[position angle, $PA$= 193$\degr$, as reported by][]{Baines06}, and then rotated by 180$\degr$. Gaussians were fitted along the spatial direction in each individual bias-subtracted, flat-fielded spectrum from the \emph{IRAF/fitprofs} routine, providing a photocenter and FWHM value for each pixel in the spectral axis. Wavelength calibration lamps were also taken and the intensity spectra were normally reduced using \emph{IRAF}. Figure \ref{Fig:gu_cma} (left) shows the intensity, photocenter, and FWHM spectra of GU CMa with the slit in the parallel direction. The two slit orientations rotated by 180$\degr$ serve to address possible instrumental artifacts; real photocenter shifts appear reversed, whereas instrumental effects and FWHM signals remain the same in both positions \citep[see, e.g.,][]{Bailey98a,Brannigan06,Baines06}. The final photocenter and FWHM spectra are given by averaging the individual spectra in both orientations. The continuum-corrected photocenter spectrum can then be computed by dividing the observed spectrum by a wavelength-dependent factor I$_{\lambda}$/(I$_{\lambda}$ +1) \citep[see, e.g.,][]{Whelan08}. Similarly, the slit was aligned in two orientations along the perpendicular direction (103$\degr$ and 283$\degr$). Figure \ref{Fig:gu_cma} (right) shows the corresponding averaged intensity, photocenter, and FWHM spectra. The continuum-corrected photocenter spectrum is not plotted in this case as it is practically the same as the averaged one.

The small difference in the H$\alpha$ peak intensity for the two slit positions is due to short-term line variability \citep[see, e.g.,][]{Praderie91}. The expected photocenter shift in the parallel direction is $\sim$ 193 mas, assuming c$_{H\alpha}$ $\sim$ 0.95 magnitudes \citep{Pogodin11} and $s$ $\sim$ 660 mas \citep{Baines06}. The continuum corrected photocenter shift in that direction is comparable ($\sim$ 189 mas) but slightly smaller than  predicted, indicating that either the contrast has changed and/or the companion has displaced from the position reported by \citet{Baines06} more than a decade ago. Indeed, there is also a marginal signal of $\sim$ 2.3 mas  in the perpendicular position, indicating that the location of the companion has a component projected in this direction too. Remarkably, the weak photocenter signal in the perpendicular direction can be detected even when this is comparable to the noise of the adjacent continuum as given by the standard deviation ($\sim$ 2.1 mas; horizontal dashed lines in the mid-right panel of Fig. \ref{Fig:gu_cma}). The spectra in the right panels were then rebinned so that the signals can be better observed. In particular, the photocenter shift in the perpendicular direction is detected at an improved $>$ 2$\sigma$$_{phot}$ level after rebinning. The FWHM spectra decrease in H$\alpha$ with respect to the adjacent continuum at a level $\gtrsim$ 3$\sigma$$_{FWHM}$ for both slit positions, indicating a smaller line emitting region. Overall, the observed photocenter and FWHM spectra are consistent with a companion located at $PA$ $\sim$ 198$\degr$ with a similar continuum brightness than the central star, and with the emission line spectrum dominated by the latter \citep[see][for additional details]{Wheelwright10}, demonstrating the spectro-astrometric feasibility of WHT/ISIS. 

\subsection{LkCa 15}
\label{Sect:lkca15}
\begin{figure}
\centering
 \includegraphics[width=9cm,clip=true]{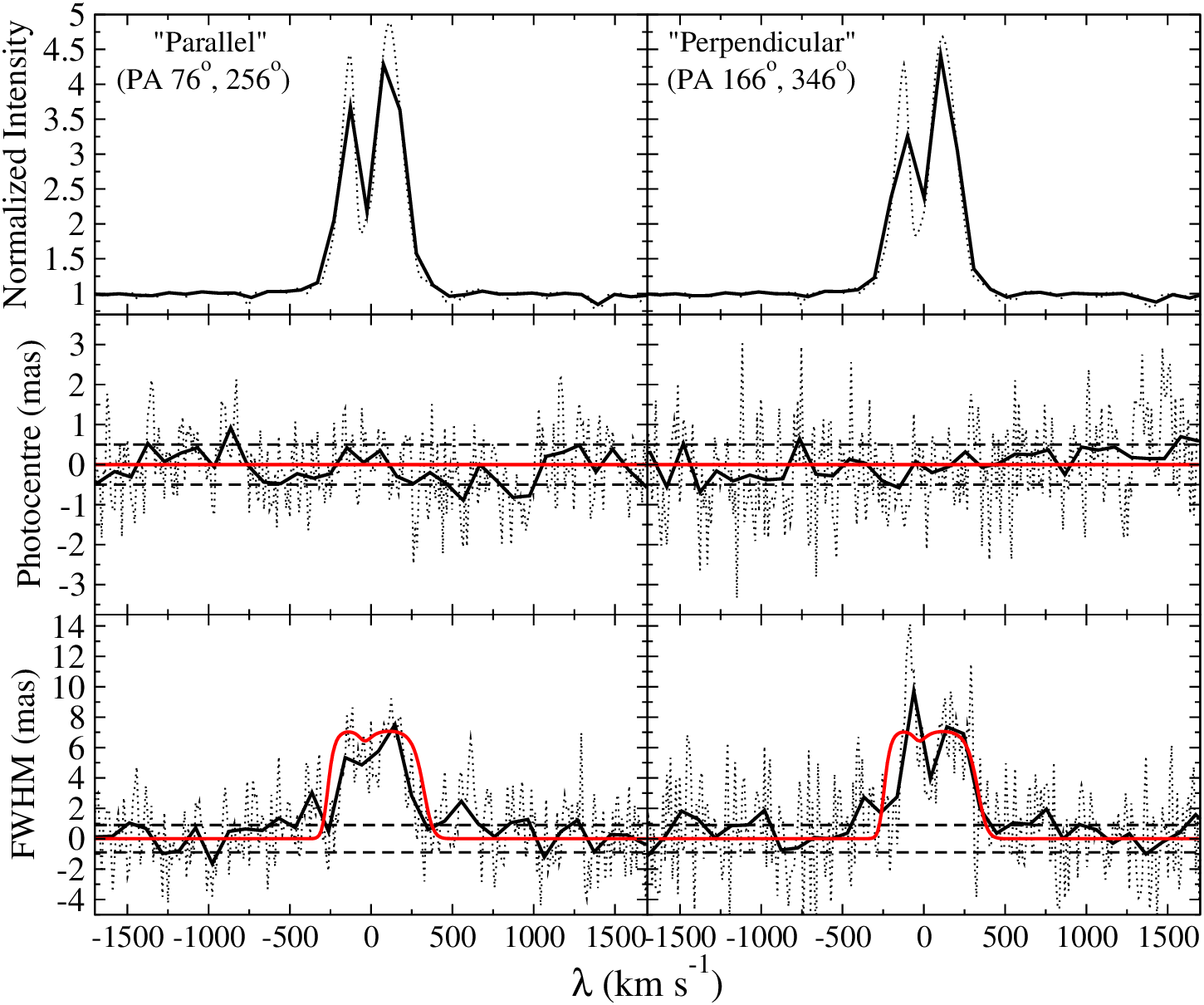}
\caption{Averaged intensity, photocenter, and FWHM spectra of LkCa 15 from top to bottom for the parallel (left) and perpendicular (right) slit orientations from position angles as indicated in the legend, both with the original spectral resolution (black dotted lines) and after rebinning (black solid lines). The horizontal dashed lines in the middle and bottom panels indicate the $\pm $1$\sigma$ levels for the rebinned spectra, corresponding to $\pm$ 0.5 and $\pm$ 0.9 mas for the photocenter and FWHM accuracies. The red lines shows a face-on disk/sphere model with inverse Gaussian profile and radius $\sim$ 70 mas or $\sim$ 11 au at 159 pc (see Sect. \ref{Sect:extended emission} and Appendix \ref{Sect:appendixb}).}
\label{Fig:lkca15}
\end{figure}

Figure \ref{Fig:lkca15} shows the spectro-astrometric results for LkCa 15. Data were reduced as for GU CMa, although a total of 16 individual spectra with typical exposure times of 15 min each were taken in the case of LkCa 15 (Table \ref{table:spectra}). These were weighted by the corresponding signal-to-noise ratios ($S/N$s) before averaging to derive the final intensity, photocenter, and FWHM spectra. The continuum-corrected photocenter spectra do not show significant differences with respect to the originals, for which these are kept. The parallel direction refers to $PA$ = 256$\degr$ (and also 256 - 180 =76$\degr$), as reported by \citet{Sallum15} for the candidate b emitting in H$\alpha$. The spectra were also binned in the spectral axis in order to provide the best photocenter and FWHM accuracy without losing the shape of the double-peaked H$\alpha$ intensity profile. The wavelength axis was converted into velocity units by applying radial velocity correction to several photospheric lines.  

The intensity spectrum is variable on timescales of less than an hour \citep[see, e.g.,][Sect. \ref{Sect:outflow}, and Appendix \ref{Sect:appendix-1}]{Bouvier93}, which explains why  the average intensity profiles are slightly different in the parallel and perpendicular slit directions. No photocenter signal can be measured at either direction up to a maximum accuracy of $\pm$ 0.5mas (=$\sigma$$_{phot}$). In contrast, a change in the FWHM is observed in the parallel and perpendicular directions at a $>$3$\sigma$$_{FWHM}$ level ($\sigma$$_{FWHM}$ = 0.9mas). Appendix \ref{Sect:appendix0} shows that the origin of such a FWHM signature is neither instrumental nor related to the data reduction process. However, it is not clear whether the minor differences between the FWHM signals in the parallel and perpendicular positions (it is slightly larger in the latter) are real or not, given that such differences are comparable to the noise. The interpretation of the spectro-astrometric results of LkCa 15 is not straightforward and  are analyzed in the next section.

\section{Analysis and discussion}
\label{Sect:analysis}
\subsection{Inconsistency with an accreting planet}
\label{Sect:planet}
The first question is whether our spectro-astrometric observations of LkCa 15 are consistent with a star + planet system with the properties reported in \citet{Sallum15}. The answer is negative, which is justified next and discussed in detail in Appendix \ref{Sect:appendixa}. As  was introduced in Sect. \ref{Sect:Intro}, such a planet should cause a weak photocenter shift of $\sim$ 0.8 mas in H$\alpha$ at the parallel slit position. In addition, because both the planet and the star would emit in H$\alpha$ \citep[the star shows near-UV excess consistent with accretion at $\sim$ 4 $\times$ 10$^{-9}$ M$_{\odot}$ yr$^{-1}$;][]{Manara14}, the size of the line emitting region and therefore the FWHM should also increase at this slit direction. However, this FWHM signature should be an order of magnitude smaller than our detection limits according to the simulations in Appendix \ref{Sect:appendixa}, which also confirms the rather obvious conclusion that no photocenter or FWHM signals should be observed when the slit is oriented in the perpendicular position. Therefore the model predictions do not fit our observations given that, on the contrary, they reveal no photocenter shift at any slit position, and a FWHM signal at both slit positions. Appendix \ref{Sect:appendixa} shows that the presence of a planet could still be consistent with the lack of photocenter signature if it has an H$\alpha$ contrast $\gtrsim$ 5.5 magnitudes with respect to the central star or $\lesssim$ 6 magnitudes in the adjacent continuum. In either case, the FWHM signals cannot be explained by such a planet. 

An alternative is that the planet has moved around the central star during the $\sim$ 3 years elapsed between the observations in \citet{Sallum15} and our recent spectra. Based on the orbit predicted in that paper and our modeling in Appendix \ref{Sect:appendixa}, the actual position of the planet should provide a photocenter shift in our now ``close to parallel'' slit position that would be even easier to detect than in the previous case,  accompanied by a FWHM signal below detection limits. Although a component should now be projected onto the ``close to perpendicular'' slit position too, the corresponding photocenter and FWHM signals would both be below detection limits for this slit orientation. Therefore, our observations are not consistent with the planet reported in \citet{Sallum15} neither for its past or its current position. Moreover, a planet causing a FWHM signal similar at both slit orientations should be located at $\sim$ 45$\degr$ from them, meaning that its orbit is far from that predicted. In fact, our observations cannot be reproduced by a star+planet system for any combination of separation, brightness contrast and planet position, the reason being that if a planet is causing the observed FWHM signature at both slit positions then a much larger photocenter shift should be observed too \citep[see][and Appendix \ref{Sect:appendixa}]{Baines06}.  

\subsection{Extended and roughly symmetric H$\alpha$ emission}
\label{Sect:extended emission}
The second question is whether any system can explain our spectro-astrometric observations. All possibilities that fit our data fall within the general category of an extended H$\alpha$ emitting region that appears roughly symmetric on the sky. Such a structure would naturally explain the lack of signature in the position spectrum (the photocenter of the system coincides with the position of the central star at all wavelengths) and the similar FWHM signal observed at both slit positions (the extent of the line emitting region is larger than that of the continuum and comparable in all directions). Appendix \ref{Sect:appendixb} shows spectro-astrometric simulations for a face-on disk. A spherical structure looks the same projected in the sky and both are equivalent for this discussion. Four different scenarios that depend on the distribution of the H$\alpha$ emission were considered, providing face-on disk/sphere line emitting regions with an extent between $\sim$ 11 and 26 au at a distance of 159 pc (see Table \ref{table:fit}). 

Inclined disk models are also considered in Appendix \ref{Sect:appendixb} for different inclinations and $PA$s of the major axis. The recent results on the geometry of the inner dust disk are used as a guide \citep[$i$ $\sim$ 45$\degr$, $PA$ $\sim$ 70$\degr$;][]{Oh16}. However, the previous values for $i$ and $PA$ do not fit our observations for any value of the gas disk radius, providing a FWHM signal for the slit orientation close to the disk major axis ($PA$ = 76$\degr$) significantly larger than for the perpendicular direction. Assuming that the inner gas disk inclination is $\sim$ 45$\degr$, a relatively small range in $PA$, roughly between 10 and 35$\degr$, projects similar disk sizes onto our slit orientations and is consistent with our observations. Alternatively, the $PA$ can be fixed to 70$\degr$, but then the H$\alpha$ disk should be close to pole-on ($i$ $<$ 35$\degr$) in order to probe similar spatial scales with our slit orientations and fit the observed FWHM signatures. All possible inclined disk configurations have gas disk radii equal to or slightly larger than the above-mentioned radii for the modeled face-on disks/spheres, but significantly smaller than the dust disk size \citep[$\sim$ 35 au at 159 pc,][]{Oh16}. Therefore, the radius of the smallest face-on disk/sphere model ($\sim$ 11 au) is a lower limit for the size of the H$\alpha$ emitting region and is plotted with red lines in Fig. \ref{Fig:lkca15}. An H$\alpha$ gas disk model with its center displaced from the star by a similar amount than the dust disk \citep{Thalmann15,Thalmann16,Oh16} was also considered in Appendix \ref{Sect:appendixb}, but such a misalignment should cause a measurable photocenter signal that we do not observe. 

\subsection{Possible relation with a variable disk wind}
\label{Sect:outflow}
Our photocenter and FWHM spectro-astrometric observations cannot be explained by an orbiting forming planet but by  a roughly symmetric H$\alpha$ emission extended $\sim$ 10--30 au and centered on the star. Some constraints on the possible physical origin of such H$\alpha$ emission are discussed in the following. 

First, an atomic gas disk rotating Keplerian has been observed in other pre-transitional stars \citep[e.g., HD 100546 in][]{Mendi15}. However, such a disk should cause a specific signature in the position spectra given that the photocenter of the disk region approaching  us would be blueshifted from the continuum photocenter of the central star and from the redshifted component of the disk \citep[see, e.g.,][and references therein]{Wheelwright12,Brittain15}. Moreover, if the two peaks observed in the H$\alpha$ intensity profile of LkCa 15 at $>$ 100 km s$^{-1}$ represent Keplerian velocities, the corresponding radial distances for all disk inclinations would be at least one order of magnitude smaller than inferred from the observed FWHM signal. Therefore, if we assume that the H$\alpha$ emitting region has the shape of a disk and this disk is not face-on, the atomic gas is most probably not moving according to  Keplerian rotation but moving radially with a comparatively negligible azimuthal component. 

Another common explanation for non-stellar H$\alpha$ emission in pre-main sequence (PMS) stars is magnetospheric accretion \citep[e.g.,][]{Muzerolle98,Kurosawa06,Mendi17a}. This scenario predicts that the emitting gas comes from the magnetic streams launched from the disk truncation radius, which for the $\sim$ kG magnetic field of LkCa 15 corresponds to R$_t$ $\sim$ 8.35R$_*$ \citep{Manara14}, i.e., 2 orders of magnitude below the estimated line emitting size. Moreover, gas is accreted magnetospherically if R$_t$ $<$ R$_{cor}$, being R$_{cor}$ = (GM$_*$R$_*^2$/v$_*^2$)$^{1/3}$ the corotation radius. Otherwise if R$_t$ $>$ R$_{cor}$ the gas is ejected as a wind from a wider range of disk radii \citep{Shu94}. The stellar parameters in \citet{Manara14} and a rotational period of $\sim$ 5.9 days from \citet{Bouvier93} are consistent with a R$_{cor}$ comparable to R$_t$, for which a wind contribution cannot be neglected to explain the H$\alpha$ emission in LkCa 15. In fact, its spectrum shows the forbidden \ion{[O]}{I} line at 6300 $\AA$ with both a blueshifted low-velocity ($\sim$ -9 km s$^{-1}$) and high-velocity ($\sim$ -90 km s$^{-1}$) components \citep{Manara14}, which is a clear indication of disk winds and jets \citep[e.g.,][]{Kwan88,Hartigan95,Rigliaco13} or perhaps a photoevaporative outflow \citep{Ercolano10}, given the strong X-ray emission detected in this source \citep{Skinner17}. However, additional forbidden lines like \ion{[N]}{II} 6548, 6583 and \ion{[S]}{II} 6716, 6731, indicative of high-velocity, low-density jets and outflows perpendicular to the disk \citep[e.g.,][]{Hartigan95,Whelan04} are not apparent in our optical spectra of LkCa 15 (Fig. \ref{Fig:fullspec}). The presence of \ion{[O]}{I} 6300 and the absence of the rest of the forbidden lines would be consistent with a slower, higher density disk wind \citep[][and references therein]{Rigliaco13}. Indeed, an extended disk wind with a relatively slow projected radial velocity comparable to our spectral resolution ($\sim$ 30 km s$^{-1}$) would explain the observed FWHM signal without any photocenter shift in our spectro-astrometric observations. Extended disk winds of several dozens of au have been reported before for other PMS stars \citep[e.g.,][]{Coffey08,Sacco12,Klaasen13,Bjerkeli16}.

Regarding variability, Appendix \ref{Sect:appendix-1} suggests that the data corresponding to our spectro-astrometric analysis, where the H$\alpha$ profile has a blue peak fainter than the red and a blueshifted central dip, is not typical for LkCa 15 (Fig. \ref{Fig:var}). Remarkably, \citet{Kurosawa06} shows that such an H$\alpha$ profile can be reproduced from an extended disk wind (see, e.g., the case $\beta$ =2, T = 7000 K in their Fig. 9), whereas a more compact accretion contribution is needed to reproduce an H$\alpha$ profile with the blue peak considerably brighter than the red \citep{Muzerolle98,Kurosawa06}, as it seems more usual for LkCa 15. Figure 2 in \citet{Whelan15} actually shows that when the H$\alpha$ red and blue peaks become comparable and the central dip changes from redshifted to blueshifted, the highly redshifted absorption apparent in other lines shown by LkCa 15 (H$\gamma$, HeI10830, Pa$\beta$, etc.), suggestive of accretion, vanishes as well. Given the strong variability it is reasonable to think that the main contributor to the H$\alpha$ emission and the size of the emitting region could also change, and our spectro-astrometric data could be reflecting a stage where an extended wind could dominate. Still, a robust physical explanation requires specific radiative transfer modeling and would benefit from multi-epoch, multi-wavelength spectro-astrometry at high resolution and high $S/N$.

Finally, the line emitting size inferred from our observations is comparable to the radial distance of the H$\alpha$ emission linked to LkCa 15 b by \citet{Sallum15}. An extended inner disk composed of dust plus atomic gas of variable size might account not only for the near-IR signal attributed to the three protoplanets \citep[][]{Thalmann16}, but could also be the origin of the H$\alpha$ emission originally associated with LkCa 15 b.

\section{Summary and conclusions}
\label{Sect:conclusions}
Our spectro-astrometric observations of LkCa 15 cannot be reproduced by an orbiting accreting planet emitting in H$\alpha$, but can be explained by a roughly symmetrical emitting region extended several au, perhaps related to a variable disk wind.  

We have tested the presence of a planet with an H$\alpha$ brightness contrast of $>$ 5 magnitudes at a radial distance $<$ 100 mas from a relatively faint central star with R $>$ 11 mag, using a medium-sized telescope for a total observing time $<$ 5 hours. These numbers can be significantly improved by using larger Earth-based telescopes assisted by adaptive optics or space telescopes unaffected by seeing, or by observing brighter sources. This, along with the fact that spectro-astrometry is able to recover the emission line spectrum of the companion from the photocenter shift, makes this technique ideal for future surveys whose aim is to detect accreting planets around young stars and to understand the way they grow. Therefore, spectro-astrometry complements differential polarimetry, high-resolution imaging, interferometry, or sparse aperture masking, with advantages like the high sensitivity to faint sources, the relatively simple data reduction and interpretation, and the use of comparatively cheap instrumentation like a long-slit spectrograph.

\begin{acknowledgements}
The authors thank the anonymous referee for the comments, which  helped to improve the manuscript. Based on observations made with the WHT and INT operated on the island of La Palma by the Isaac Newton Group of Telescopes in the Spanish Observatorio del Roque de los Muchachos of the Instituto de Astrofísica de Canarias. This work has made use of data from the European Space Agency (ESA) mission {\it Gaia} (\url{https://www.cosmos.esa.int/gaia}), processed by the {\it Gaia} Data Processing and Analysis Consortium (DPAC \url{https://www.cosmos.esa.int/web/gaia/dpac/consortium}). Funding for the DPAC has been provided by national institutions, in particular the institutions participating in the {\it Gaia} Multilateral Agreement. IM acknowledges the Government of Comunidad Aut\'onoma de Madrid, Spain, which has funded this work through a `Talento' Fellowship (2016-T1/TIC-1890)

\end{acknowledgements}

\bibliographystyle{aa} 
\bibliography{myrefs.bib} 

\begin{appendix} 

\section{H$\alpha$ variability of LkCa 15}
\label{Sect:appendix-1}
\begin{figure}
\centering
 \includegraphics[width=9cm,clip=true]{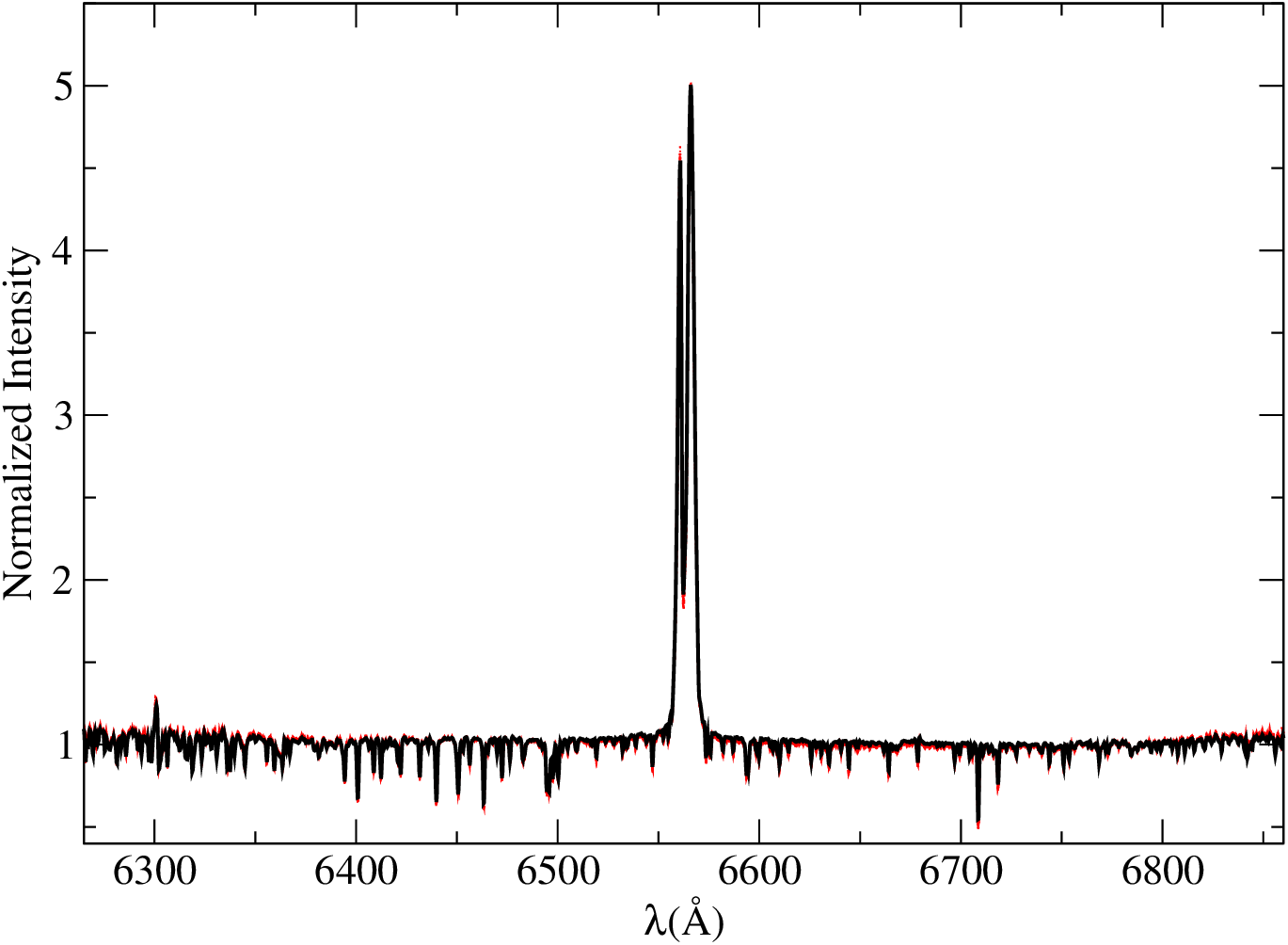}
\caption{Multi-epoch optical spectra of LkCa 15 taken with ISIS/WHT (red dotted lines) and average (black solid line).}
\label{Fig:fullspec}
\end{figure}

Figure \ref{Fig:fullspec} shows the full optical spectra of LkCa 15 as obtained with ISIS/WHT. Apart from H$\alpha$, the only emission line that is apparent in the spectra is the forbidden \ion{[O]}{I} 6300. However, this lies close to the edges of the detector and the corresponding $S/N$ is not high enough for accurate spectro-astrometric analysis. 

\begin{figure}
\centering
 \includegraphics[width=9cm,clip=true]{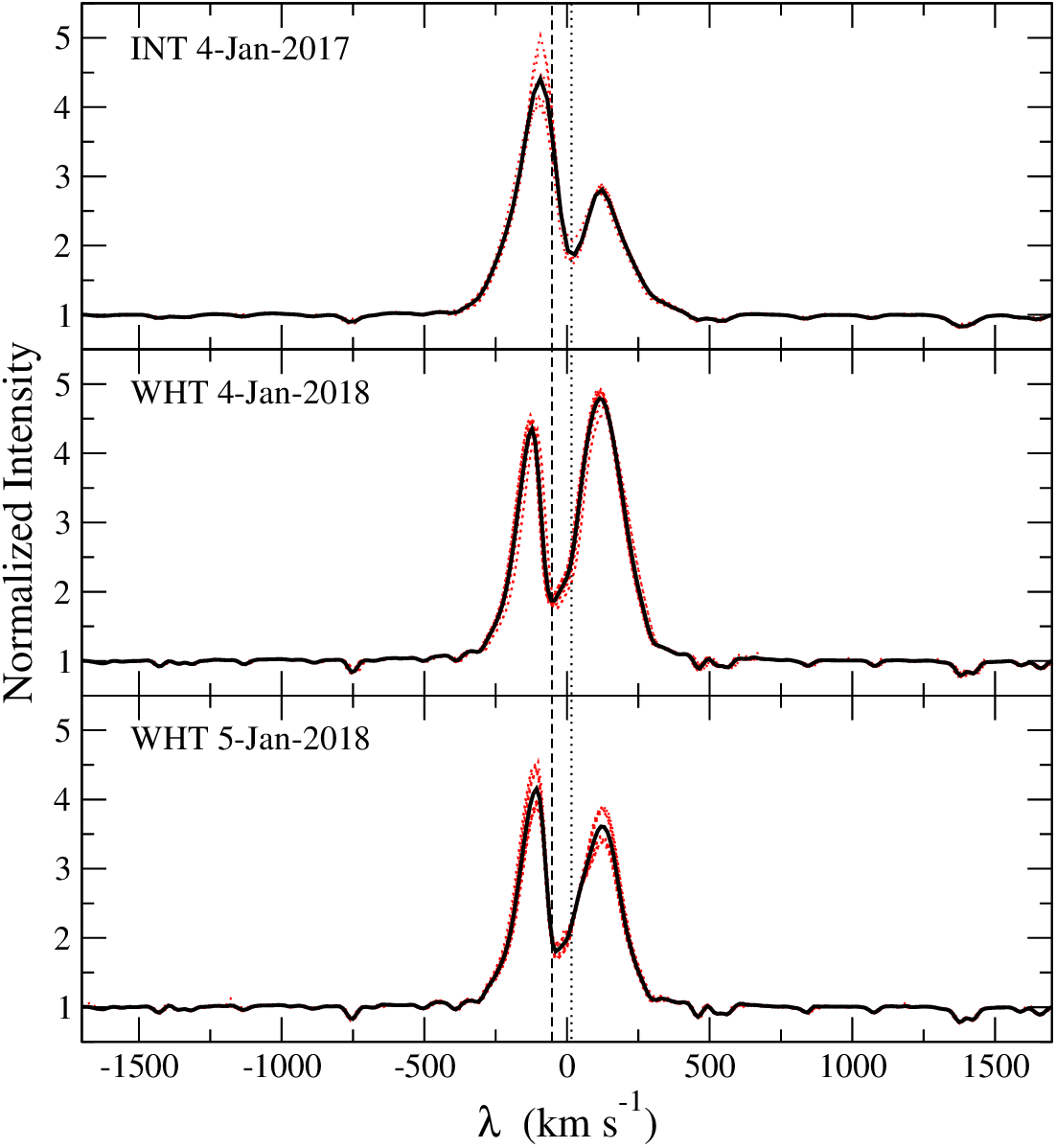}
\caption{Multi-epoch H$\alpha$ spectra of LkCa 15 taken with the telescopes indicated. The red dotted lines are the individual spectra and the black solid lines the corresponding averages. The vertical dashed and dotted lines indicate the position of the central dip in the middle ($\sim$ -50 km s$^{-1}$) and upper ($\sim$ +15 km s$^{-1}$) panels, respectively. The spectro-astrometric analysis in this work corresponds to the data in the middle panel.}
\label{Fig:var}
\end{figure}

Figure \ref{Fig:var} shows several H$\alpha$ profiles of LkCa 15, illustrating part of its variability. The spectra in the top panel were taken in service mode with the IDS spectrograph at the Isaac Newton Telescope (INT) in La Palma Observatory as part of our spectro-astrometric campaigns. The rest of the spectra correspond to the ISIS/WHT data analyzed in this work (middle panel) and additional data that we took during the first half of the following night (bottom panel). Unfortunately, the weather conditions and the $S/N$ of the spectra were good enough to carry out spectro-astrometric analysis only during the night of 4 January 2018, as presented in this letter. The intensity spectra in Fig. \ref{Fig:var} and additional spectra taken with CAFE at the 2.2m telescope in Calar Alto Observatory and XShooter/VLT on 2011 and 2012 \citep[Emma Whelan, private communication, and][respectively]{Whelan15} suggest that the H$\alpha$ profile of LkCa 15 is typically double peaked with the blue peak brighter than the red, which is also observed in additional archival spectra taken with FEROS at the 2.2m MPG/ESO telescope on 2013. In contrast, the H$\alpha$ profiles corresponding to our spectro-astrometric analysis show the blue peak fainter than the red, but we note that the relative brightness of the two peaks changed again the following night. Regarding the central dip, this seems to be usually centered on the star or redshifted, which suggests accretion, whereas in our ISIS/WHT campaign it was blueshifted, which suggests outflows/winds.    

Although other H$\alpha$ line profiles or even an absence of emission have  been reported in the past \citep[e.g.,][and references therein]{Bouvier93}, the data mentioned above suggest that the spectro-astrometric observations analyzed in this work could be reflecting an H$\alpha$ line profile that is not usual for LkCa 15, at least in  recent years. We suggest that a detailed variability study could be very useful to constrain the physical origin of the H$\alpha$ emission of LkCa 15.

\section{ FWHM signature in LkCa 15}
\label{Sect:appendix0}
Given the relevance of the FWHM signature detected in the spectro-astrometric analysis of LkCa 15, we carried out the following tests in order to understand whether it could be associated with instrumental, data reduction, or methodological effects.

\begin{itemize}
 \item Based on our complementary IDS/INT data with a number of counts that exceeds the saturation limit of the CCD, we conclude that artificial FWHM signatures can arise in overexposed spectra. However, such artificial FWHM signals are sharp and narrow, clearly contrasting with the weak and broad signature detected with ISIS/WHT. Table \ref{table:spectra} summarizes the properties of each ISIS/WHT spectrum that has been analyzed in this work to derive the spectro-astrometric observables. Peak counts were below the saturation limit in all cases (the nominal value is 65535 counts), which along with the shape of the FWHM signature indicates that this is not caused by CCD saturation problems.
 \item As described for GU CMa in Sect. \ref{Sect:gucma}, the use of complementary spectra rotated by 180$\degr$ and the fact that the FWHM signals are only observed in H$\alpha$ provide additional confidence that these signatures are not due to pixel problems or instrumental artifacts.
 \item The FWHM change is also apparent in the individual raw spectra of LkCa 15 before subtracting the bias and dividing by the flatfields, guaranteeing that data reduction is not causing the appearance of an artificial signature. The bias/flatfield correction has only the expected effect of reducing the noise.  
 \item The \emph{IRAF/fitprofs} routine was applied several times with distinct input parameters aiming to remove the FWHM signature, but without success. In particular, the use of different background constraints does not have any effect on the observed FWHM change. We also built an alternative MATLAB routine to fit the Gaussian center and FWHM to the spatial profiles of the 2D spectra, finding the same FWHM spectra and discarding a methodological effect.  
 \item Finally, we  checked that the Poisson nature of the data and the sampling of the cross-dispersion profile at discrete points do not introduce effects comparable to the observed FWHM. In particular, we set up a simulation that models the cross-dispersion taking into account the finite size of the pixel and simulated a cross-dispersion profile following the Poisson statistics. Running 10$^6$ independent simulations for a variety of amplitudes,  for the parameters at hand we found
 no significant trend of the FWHM with the input signal. Only at lower count numbers did we observe a slight trend of the fitted FWHM with the signal strength that likely stems from the \emph{IRAF/fitprofs} approach of minimizing $\chi$$^2$, which does not take the Poisson nature properly into account.
\end{itemize}

In summary, to the best of our knowledge the FWHM signature observed in the ISIS/WHT data cannot be attributed to any artificial effect;  in this work it has been assumed that such a feature has a physical origin associated with the H$\alpha$ line emitting region in LkCa 15.

\begin{table}
\caption{Properties of the ISIS/WHT spectra}              
\label{table:spectra}      
\centering 
\setlength{\tabcolsep}{4.5pt}
\begin{tabular}{l c c c c c c}         
\hline\hline                        
ID & JD & $PA$       & t$_{exp}$ & Peak counts & $S/N$ & $seeing$\\
...         &(+2458123) & ($\degr$)& (seg)     &...          &...& ($\arcsec$)\\
\hline                                  
$\#$1    &0.331& 76       &900           &37400&220&1.32\\
$\#$2    &0.344& 76       &1000          &40000&380&1.32\\
$\#$3    &0.356& 76       &1000          &44000&390&1.23\\
$\#$4    &0.374& 256      &1000          &63446&320&1.07\\
$\#$5    &0.389& 256      &900           &62045&270&1.01\\
$\#$6    &0.405& 256      &720           &53500&340&0.97\\
$\#$7    &0.414& 256      &720           &41061&305&1.12\\
$\#$8    &0.423& 256      &720           &41843&315&1.10\\
$\#$9    &0.432& 256      &720           &55712&300&0.94\\
$\#$10   &0.444& 166      &720           &64000&385&0.81\\
$\#$11   &0.456& 166      &570           &26550&305&1.20\\
$\#$12   &0.464& 166      &700           &34086&390&1.21\\
$\#$13   &0.475& 166      &900           &42597&335&1.24\\
$\#$14   &0.486& 166      &900           &36157&300&1.35\\
$\#$15   &0.503& 346      &1000          &44658&470&1.27\\
$\#$16   &0.515& 346      &1000          &53568&420&1.18\\

\hline                                             
\end{tabular}
\tablefoot{
Columns list spectrum identification, Julian date corresponding to 4 January 2018, slit position angle, exposure time, number of counts at the peak of the H$\alpha$ emission, H$\alpha$ $S/N$, and seeing for the ISIS/WHT spectro-astrometric observations analyzed in this work. 
}
\end{table}

\section{Spectro-astrometric modeling of a star+planet system.}
\label{Sect:appendixa} 
\begin{figure}
\centering
 \includegraphics[width=9cm,clip=true]{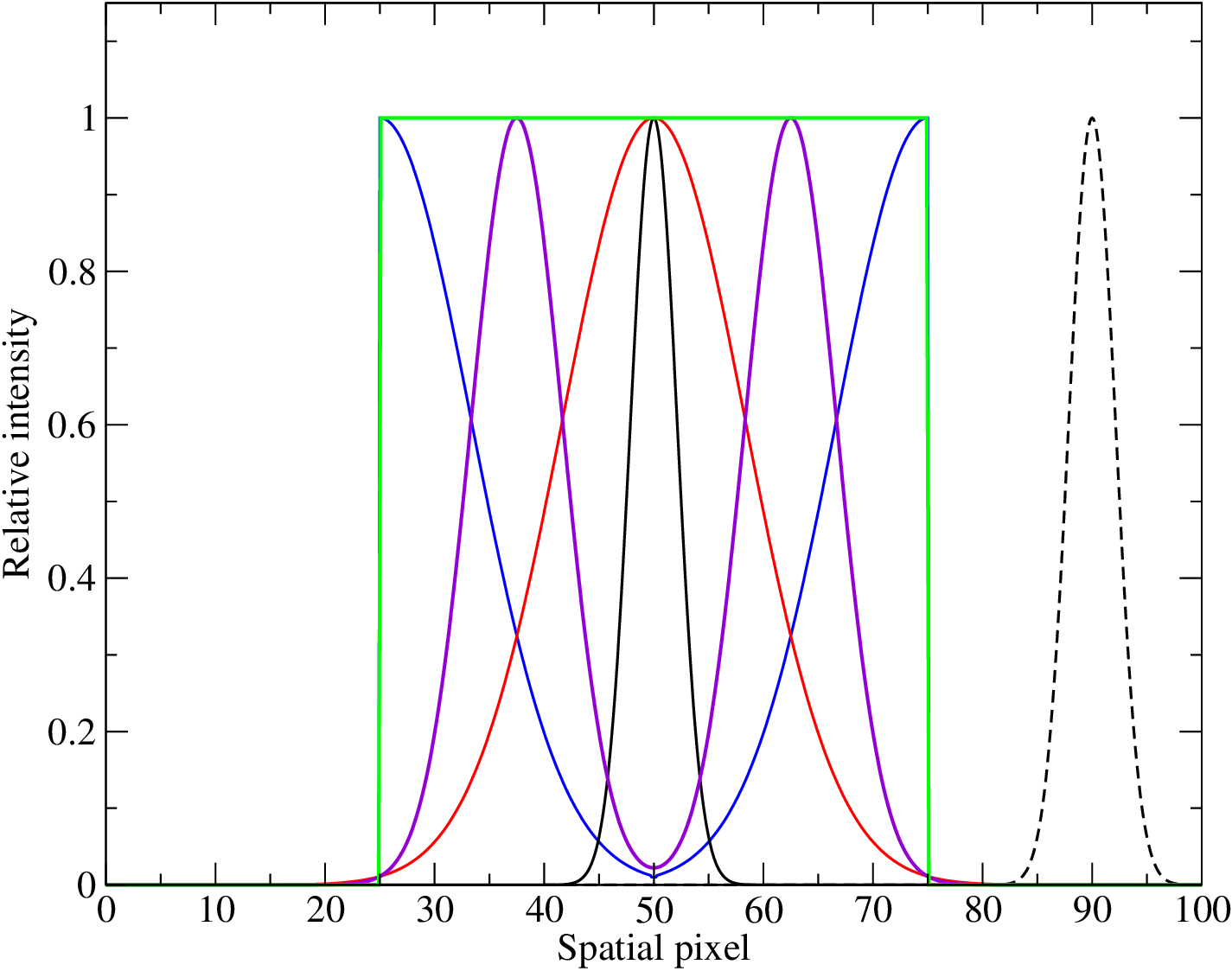}
\caption{Spatial profiles considered in the models, peaking at unity: star and planet separated by 40 pixels (solid and dashed black lines). The rest correspond to a sphere or a face-on disk with a radius of 25 pixels and whose profiles are homogeneous (green), Gaussian (red), inverse Gaussian (blue), and ring-like (violet). The star--planet separation and the sphere/disk radius are arbitrarily large compared to the ones modeled (see text), and have been selected for clarity.}
\label{Fig:profiles}
\end{figure}

We carried out spectro-astrometric modeling of a star+planet system as follows: 
\begin{itemize}
 \item The star and the planet have assumed intensities I$_*$ and I$_{planet}$ such that the sum of the two values is the total observed intensity, I$_{total}$. I$_*$ and I$_{planet}$ are also related through the brightness contrast in the H$\alpha$ emission line and the adjacent continuum, c$_{line}$ and c$_{cont}$, in magnitudes.
 \item Both the star and the planet are point sources whose corresponding spatial profiles (Fig. \ref{Fig:profiles}) are modeled as Gaussians with values weighted by I$_*$ and I$_{planet}$, a common width given by the seeing (in pixels), and means (photocenters) shifted by the on-sky separation ($s$, also in pixels). A final 2D spectrum was then computed by summing the two matrices for the star and the planet, with the values in the rows representing the spatial profiles and the values in the columns providing the spectral information.
 \item A procedure analogous to that explained in Sect. \ref{Sect:data_reduction} was finally applied to the previously derived 2D spectrum to extract the modeled photocenter and FWHM spectra, assuming that the slit is oriented at a given angle $\alpha$ with respect to the $PA$ of the planet ($\alpha$ = 0$\degr$ for the parallel position and 90$\degr$ for the perpendicular). 
 \end{itemize}
 
\begin{figure}
\centering
 \includegraphics[width=9cm,clip=true]{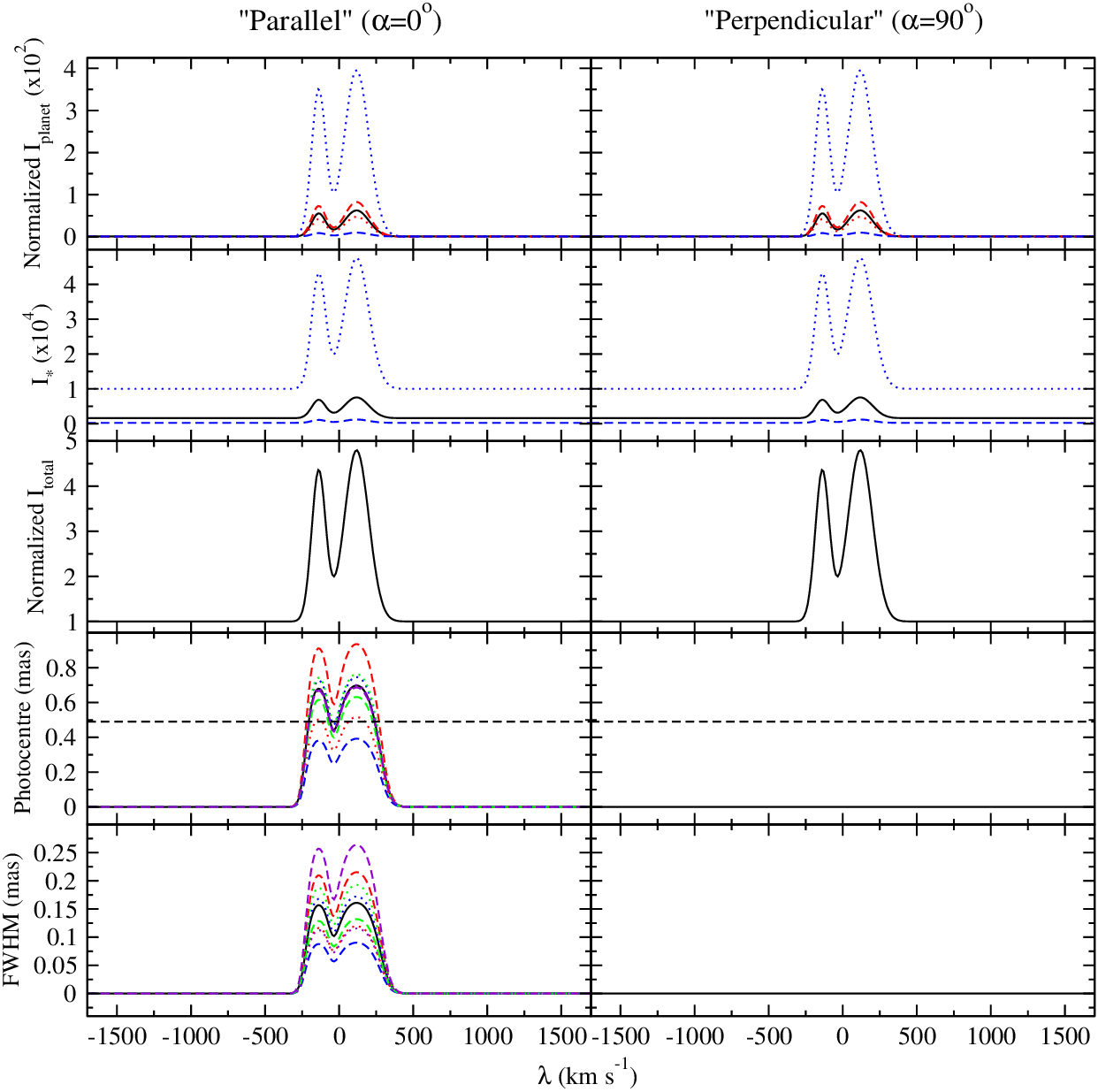}
\caption{Modeled planet, star, and total intensities, photocenter and FWHM spectra (from top to bottom) when the slit is oriented towards the planet's position and perpendicular to it (left and right panels). The black line refers to the fiducial values for LkCa 15 of the contrast in the continuum and in H$\alpha$ (5.2 and 8 magnitudes), the star--planet separation (93 mas), and the seeing (1.2$\arcsec$). The colored lines refer to variations in c$_{line}$ (red), c$_{cont}$ (blue), $s$ (green), and $seeing$ (violet) within error bars (dotted and dashed lines for the corresponding upper and lower limits; see text). The horizontal dashed lines in the photocenter panels represent our detection limits. The FWHM spectra plotted are $\sim$1 order of magnitude below detection limits.}
\label{Fig:models1}
\end{figure}

Figure \ref{Fig:models1} shows the spectro-astrometric models considering the most consistent input parameters for the LkCa 15 system and their errors \citep[for a more general approach of the spectro-astrometric signatures expected for a binary system see][]{Baines06}: c$_{line}$ = 5.2 ($\pm$ 0.3) magnitudes, c$_{cont}$= 8 ($\pm$ 2) magnitudes, and $s$ = 93 ($\pm$ 8) mas \citep{Sallum15}, which correspond to 0.42 ($\pm$ 0.04) pixels for our plate scale (Sect. \ref{Sect:data_reduction}). Similarly, $seeing$ = 5 ($\pm$ 2) pixels during our observations. The total intensity is a double-peaked H$\alpha$ profile with the blue peak fainter than the red, as observed. Given that no photocenter shift was measured, I$_*$ and I$_{planet}$ cannot be derived and are the most unconstrained input parameters of the model. Because of the large values of c$_{line}$ and c$_{cont}$, it is assumed that most of I$_{total}$ comes from I$_*$. In addition, since the width of the observed FWHM signature is comparable to that for the observed H$\alpha$ line (several hundred of km s$^{-1}$), Fig. \ref{Fig:models1} only considers the case when the H$\alpha$ line profile of the planet is equal to that for the star, which provides FWHM signals in the parallel position with a shape similar to that observed. Other planet line profiles (wide enough single Gaussians, double-peaked profiles with the blue peak stronger than the red)\ only have an effect on the shape of the photocenter and FWHM signals, but not on their strengths.

\begin{figure}
\centering
 \includegraphics[width=9cm,clip=true]{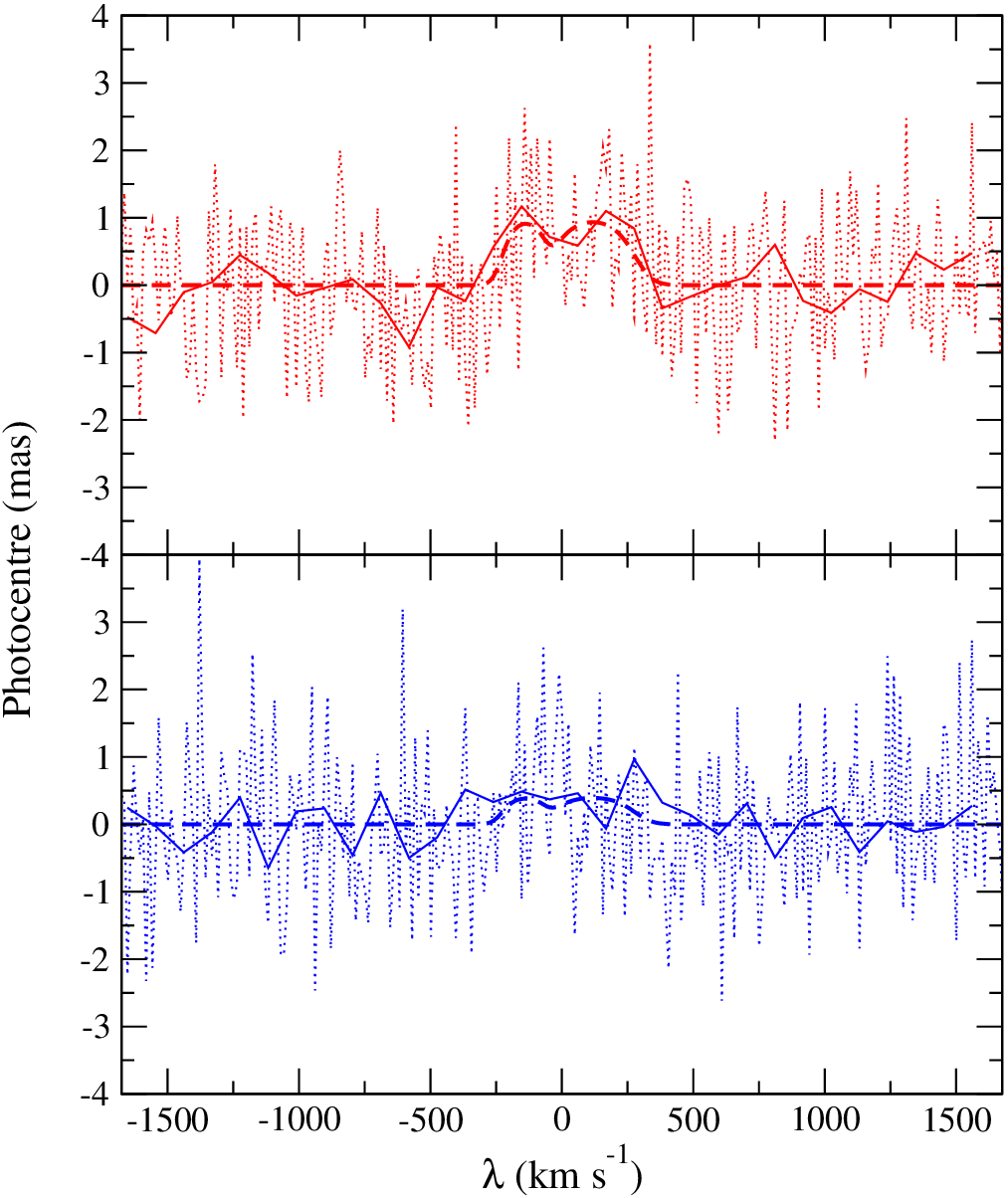}
\caption{Examples of the convolution between the maximum and minimum modeled photocenter shifts in the parallel direction from Fig. \ref{Fig:models1} (dashed lines in the top and bottom panels) with the original spectral noise and  after rebinning our ISIS/WHT spectra are plotted with dotted and solid lines, respectively.}
\label{Fig:modelsconvresolution}
\end{figure}

As  is shown in Fig \ref{Fig:models1}, an increase in c$_{line}$ within error bars, keeping the rest of the parameters unchanged, is accompanied by a decrease in the photocenter and FWHM signals in the parallel position, whereas an increase in c$_{cont}$ causes a subsequent increase in both the photocenter and FWHM. Larger star--planet  separations cause larger photocenter and FWHM signatures, and the seeing variations during our observations only have an effect on the FWHM signal, making it stronger for smaller seeing values. Remarkably, all FWHM signals predicted for the parallel position are approximately one order of magnitude below our detection limits. Regarding the modeled photocenter spectra in the parallel position, the strongest and the faintest signals in Fig. \ref{Fig:models1} were repeatedly convolved with random noise simulating that in the ISIS/WHT observed and rebinned spectra, estimating that the strongest photocenter signature should be recovered with a probability $>$ 90$\%$ from our data, but the weakest photocenter signature could only be detected at a $\sim$ 20$\%$ level. Examples of this procedure are shown in Fig. \ref{Fig:modelsconvresolution}. The detection of the planetary signal becomes more difficult for narrower H$\alpha$ profiles, and this is ultimately limited by the spectral resolution. However, we note that the underlying assumption is that the H$\alpha$ profile of the planet and the star have similar widths, as suggested by the wavelength coverage of the observed FWHM signature.

In summary, the models predict that for the reported values of LkCa15 the ISIS/WHT data should provide no FWHM signature accompanied by a photocenter shift in the parallel position above detection limits, although the lowest c$_{cont}$ (6 magnitudes) and/or largest c$_{line}$ (5.5 magnitudes) could cause undetectable photocenter shifts as well. In addition, our models confirm that no photocenter and FWHM signals should be observed for the perpendicular position. Therefore, the lack of detection in the photocenter spectra and the FWHM signals observed at both slit positions cannot be reproduced from an H$\alpha$ emitting planet as reported by \citet{Sallum15}.

\begin{figure}
\centering
 \includegraphics[width=9cm,clip=true]{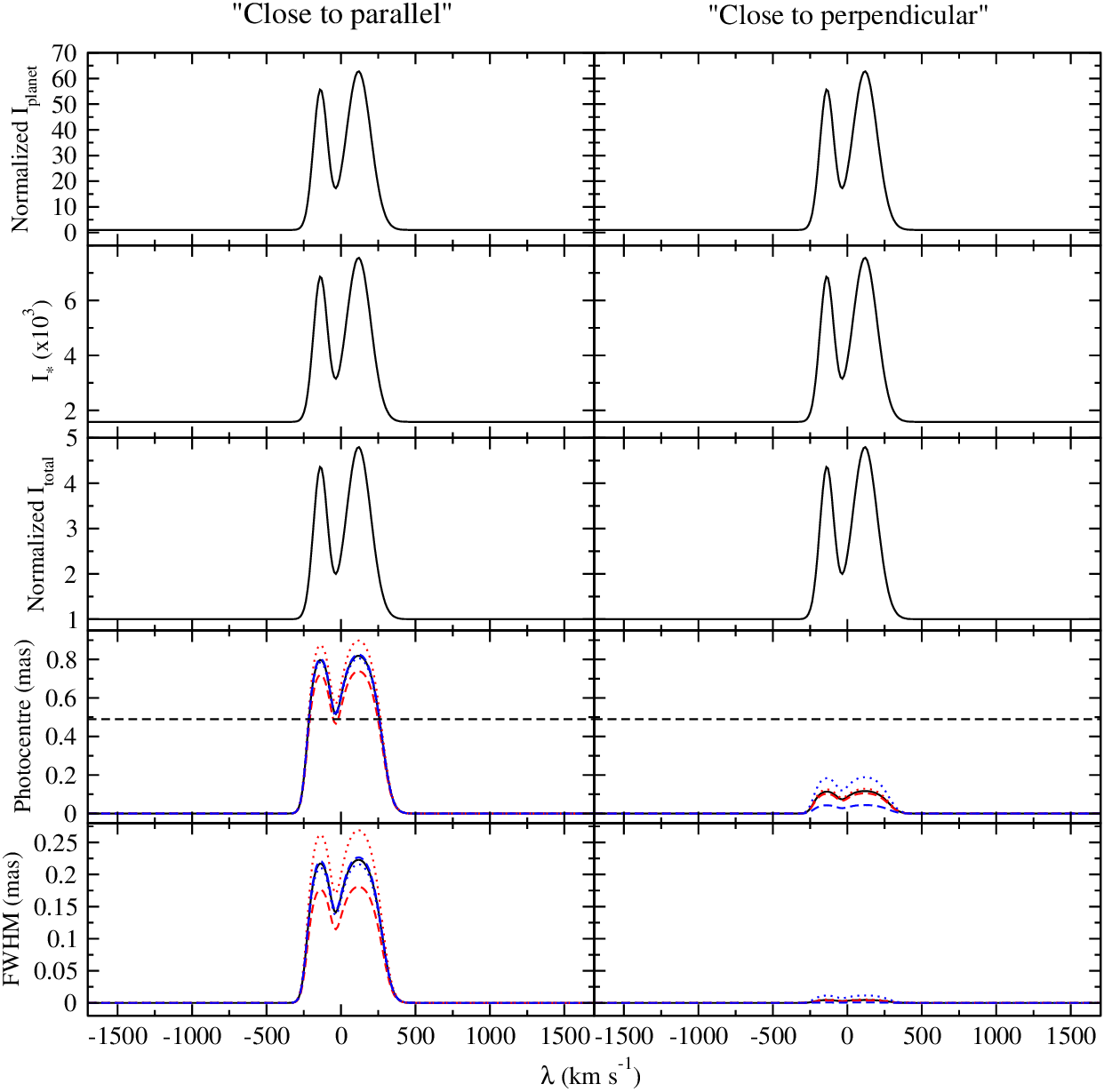}
\caption{Modeled planet, star, and total intensities, photocenter and FWHM spectra (from top to bottom) when the slit is oriented close to the planet's position and close to perpendicular (left and right panels, respectively). The black line refers to the fiducial values for LkCa 15 (see  caption in Fig. \ref{Fig:models1}), with $s$ = 110 mas, $\alpha$ = 8$\degr$ (left), and 82$\degr$ (right). The colored lines refer to variations in $s$ (red) and $\alpha$ (blue) within error bars (dotted and dashed lines for the corresponding upper and lower limits; see text).  The horizontal dashed lines in the photocenter panels represent our detection limits. The FWHM spectra plotted are $\sim$1  order of magnitude below detection limits.}
\label{Fig:models2}
\end{figure}

Also considered is the case where the candidate planet has moved from the original position reported by \citet{Sallum15}. These data were taken $\sim$ 3 years before ours, which according to the LkCa 15 b orbit prediction in that paper would mean that the current star--planet separation is $s$ $\sim$ 110 ($\pm$10) mas, i.e., 0.50 ($\pm$ 0.05) pixels. The new $PA$ is $\sim$ 248 ($\pm$ 5)$\degr$, meaning that the slit orientations used in our observing run correspond to  ``close to parallel'' and ``close to perpendicular'' positions with $\alpha$ = 8 ($\pm$ 5)$\degr$ and 82 ($\pm$ 5)$\degr$, respectively. Figure \ref{Fig:models2} shows the model results for the new $s$ and $\alpha$ values and their variations within error bars when the rest of the parameters are fixed to the fiducial values as in the previous case. The photocenter and FWHM signatures now have  components in both slit directions, although the only ones that could be measured with our instrumental configuration are once again the photocenter displacements in the close to parallel direction. Noticeably, these photocenter displacements are above detection limits in all cases, and so should be easier to measure than in the previous case. In conclusion, our observations cannot  be reconciled with an orbiting planet as described in \citet{Sallum15}.

In order to reproduce FWHM signals similar to each other in both slit directions, as observed, we need $\alpha$ $\sim$ 45$\degr$, which corresponds to $PA$ $\sim$ 211$\degr$. Not only is this value far from the orbit predicted in \citet{Sallum15}, but it also produces FWHM signals much fainter than our detection limits, as well as measurable photocenter shifts in both slit orientations. Moreover, if the strength of the FWHM signal in both slit directions is increased to a value similar to our observations by changing the values of c$_{line}$ and/or c$_{cont}$, the corresponding photocenter signatures should also increase  and would be much larger than our detection limits, in contradiction with our results. In short, we find no way to reproduce our observations from a star--planet system under any configuration.   

\section{Spectro-astrometric modeling of a star+extended emission.}
\label{Sect:appendixb}
This type of modeling was carried out in a similar way to that  for the star+planet system, with the following differences:
\begin{itemize}
 \item The H$\alpha$ line is assumed to be emitted in a circumstellar region with the shape of a sphere or a disk of a given size and center, and whose intensity is given by I$_{total}$-I$_*$. In turn, given that the photospheric absorption shown by a solar-type star like LkCa 15 is negligible compared to that observed, I$_*$ is assumed to be  continuum displaced with respect to that of the H$\alpha$ emitting material from the value c$_{cont}$.  
 \item The spatial distribution of the system in the CCD is modeled by convolving the PSF of the instrument, which is the image of the point star represented by a Gaussian with the width given by the seeing, and an assumed profile for the extended emission (Fig. \ref{Fig:profiles}), both weighted by the corresponding intensities. Therefore, the models explored here are purely geometrical without any physical assumption (see Sect. \ref{Sect:outflow}).
 \end{itemize}
The simplest scenario is when the line emitting region is a homogeneous sphere centered on the star given that this is symmetric in the sky for all slit orientations. Therefore, no photocenter shift and an equal FWHM signal are predicted for all slit positions, in agreement with our observations. Such a signal only depends on the values of c$_{cont}$ and the seeing, apart from the assumed size of the sphere. Given that the near-IR continuum emission characterizing the closest regions to the star can be represented by a blackbody at a temperature of 1600K \citep{Espaillat08}, the contrast between this and LkCa 15 \citep[4900 K][]{Manara14} is $\sim$ 10 magnitudes in the continuum close to H$\alpha$. The seeing was again set to 5 pixels. For these fiducial values, a homogeneous sphere of atomic gas with a radius in between 0.4 and 0.5 pixels fits our observations, which corresponds to 88--110 mas or 14--18 au at 159 pc. For a fixed sphere radius of 0.45 pixels, changing the contrast in the continuum by a reasonable range of $\pm$ 1 magnitude has a negligible effect of less that 0.5$\%$ on the FWHM signal. In turn, the FWHM signal decreases as the seeing increases, and a seeing range of $\pm$ 2 pixels changes the peak of the FWHM signal by $^{-2}_{+4}$ mas.

A larger sphere radius, almost twice as long as for the homogeneous sphere, is needed to fit our observations for a Gaussian spherical distribution in which the H$\alpha$ emission is distributed according to a Gaussian profile centered on the star. On the contrary, an inverse Gaussian distribution where the H$\alpha$ emission is concentrated at the edges of the sphere and smoothly decreases like a Gaussian towards the central star needs a sphere radius smaller than for the homogeneous case by a factor $\sim$ 0.8. An in-between ring-like case where the H$\alpha$ emission is concentrated at half the sphere radius from the star and decreases like a Gaussian  towards the central star and towards  the outer regions needs a radius very similar to that for the homogeneous case to fit our observations. Given the symmetry of the previous distributions, they can also be interpreted as face-on homogeneous, Gaussian, inverse Gaussian, or ring-like disks. Table \ref{table:fit} lists the sizes of the face-on disk/sphere models that fit our spectro-astrometric observations of LkCa 15, which look similar to the red line overplotted in Fig. \ref{Fig:lkca15} (inverse Gaussian case).

\begin{table}
\caption{Geometrical models}              
\label{table:fit}      
\centering                                      
\begin{tabular}{l c }         
\hline\hline                        
MODEL  & radius \\
       &(mas; au)\\
\hline                                  
    Homogeneous disk/sphere & 92; 15 \\      
    Gaussian disk/sphere & 165; 26 \\
    Inverse Gaussian disk/sphere & 70; 11 \\
    Ring disk/sphere & 103; 16 \\
\hline                                             
\end{tabular}
\tablefoot{Radii (both in mas and au at a distance of 159 pc) that fit our observations. These numbers are lower limits for the radii of the modeled inclined disks.

}
\end{table}

\begin{figure}
\centering
 \includegraphics[width=9cm,clip=true]{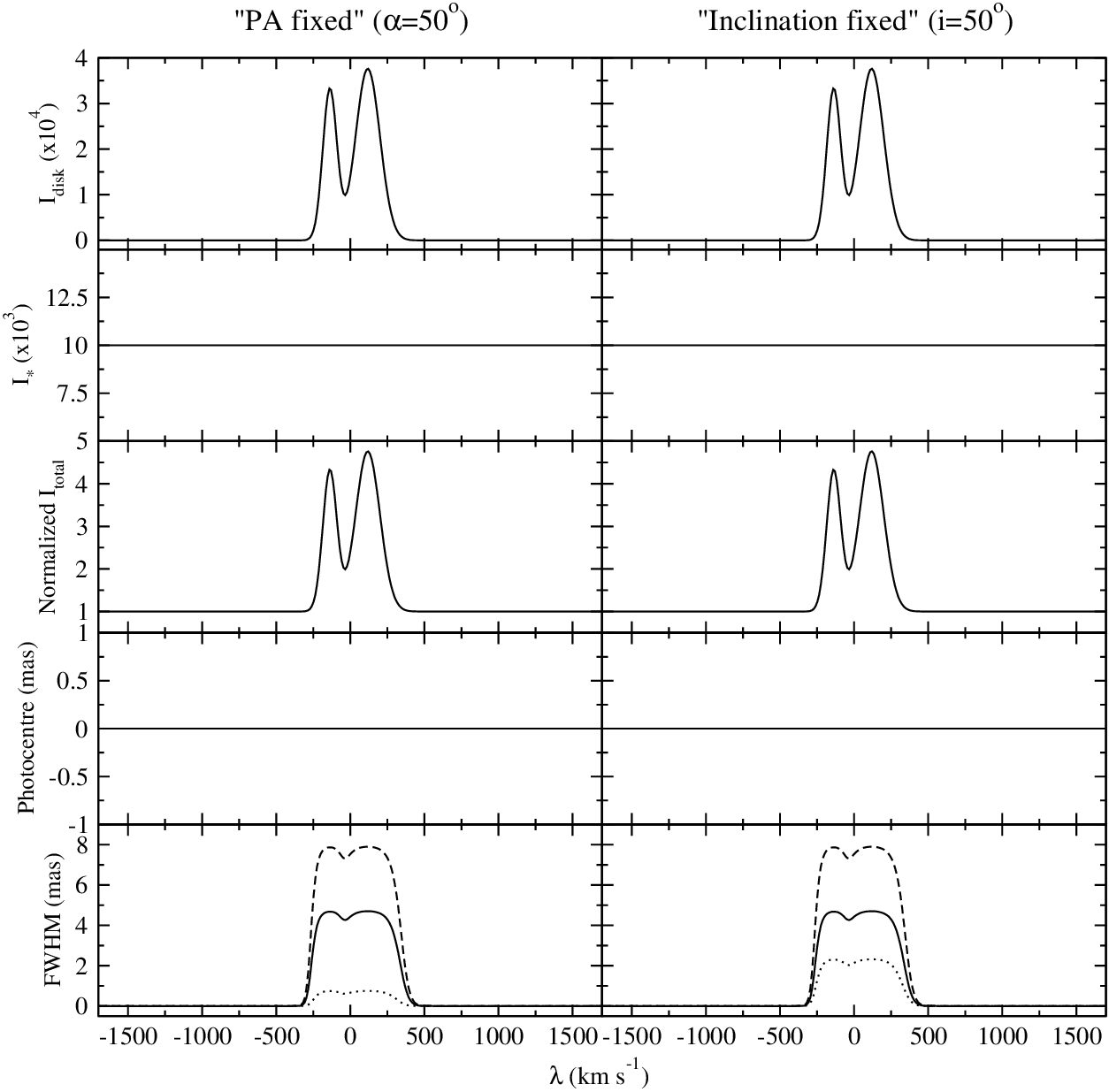}
\caption{Modeled homogeneous disk, star, and total intensities, photocenter and FWHM spectra (from top to bottom). For a fixed angle between the disk $PA$ and the slit orientation (50$\degr$), the left panels show the effect of changing the disk inclination: $i$ = 25, 50, and 75$\degr$ (dashed, solid, and dotted lines). For a fixed disk inclination (50$\degr$), the right panels show the effect of changing the angle between the disk $PA$ and the slit orientation: $\alpha$ = 25, 50, and 75$\degr$ (dashed, solid, and dotted lines). A disk radius of 0.45 pixels, brightness contrast c$_{cont}$ = 10 magnitudes and a seeing of 5 pixels is assumed in all cases.}
\label{Fig:models3}
\end{figure}

In order to model inclined disks we also need to consider the inclination to the line of sight ($i$; 0$\degr$ for face-on and 90$\degr$ for pole-on) and the $PA$ of the disk's major axis, as both parameters can have an influence on the FWHM for a given orientation of the slit; the photocenter shift always remains zero for symmetry. Figure \ref{Fig:models3} shows the case of a homogeneous disk with radius 0.45 pixels; c$_{cont}$ = 10 magnitudes; a seeing of 5 pixels; $i$ = 25, 50, and 75$\degr$; and disk $PA$s such that the angle between them and the slit's orientation is also 25, 50, and 75$\degr$. For a fixed position of the slit with respect to the disk $PA$, increasing the disk inclination causes  the FWHM signal to decrease as the apparent area of the emitting source becomes less extended. Similarly, although less acutely, the FWHM decreases as the slit is oriented at larger angles from the disk's $PA$, given that the slit probes smaller regions closer to the minor axis of the disk. The maximum FWHM signal is obtained when the slit is oriented towards the disk $PA$, for all disk inclinations. The sphere case is recovered when $i$=0$\degr$, for all slit orientations. In general, inclined disk models need larger radii to reproduce our observations than the face-on disks/sphere cases described above. 

\begin{figure}
\centering
 \includegraphics[width=9cm,clip=true]{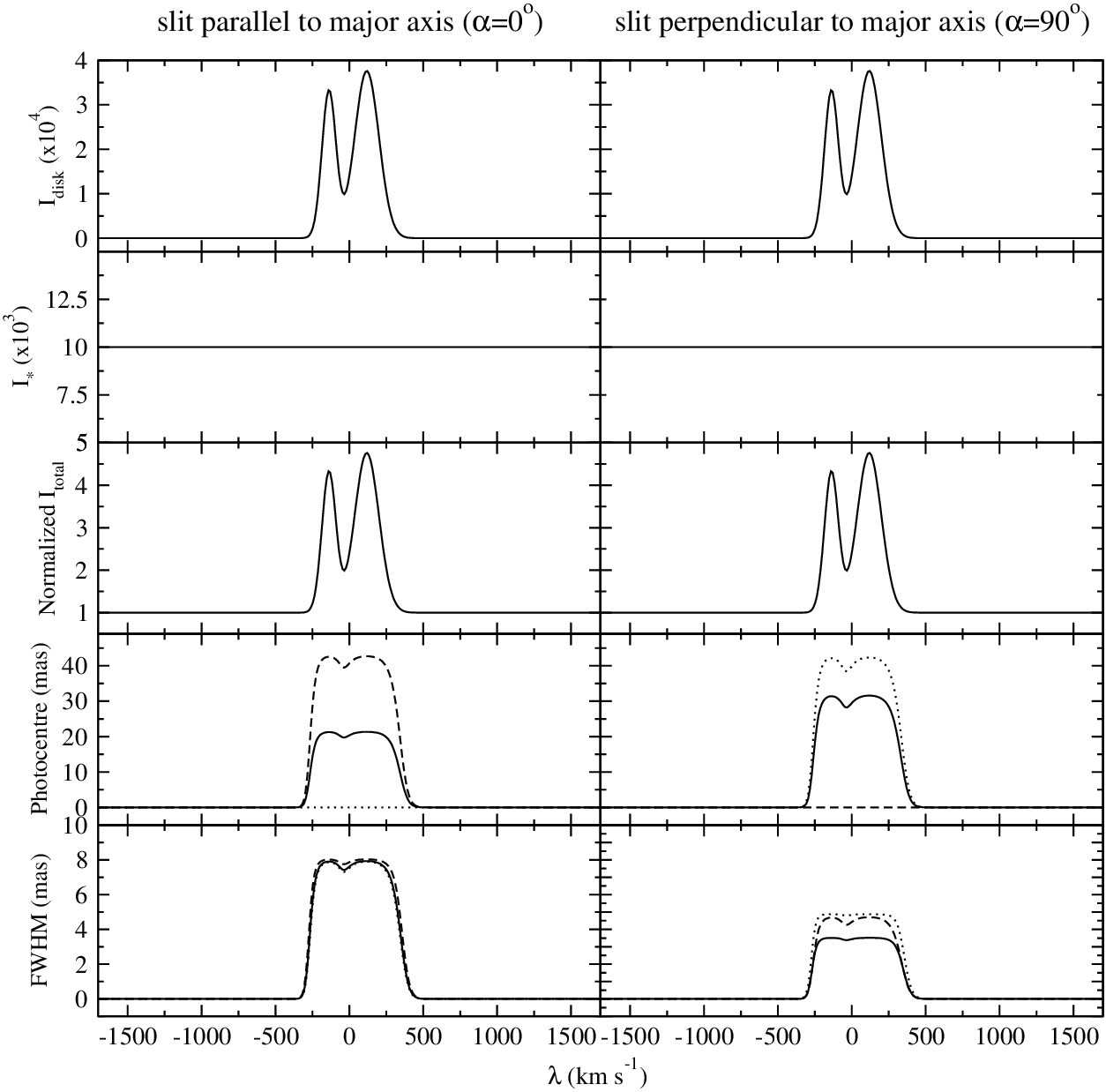}
\caption{Modeled homogeneous disk, star, and total intensities, photocenter and FWHM spectra (from top to bottom) when the center of the disk and the star are separated by 0.2 pixel and the slit is oriented towards the major and minor axes of the disk (left and right panels). The effect of shifting the disk center along the disk major axis, minor axis, and 45$\degr$ from both are indicated with the dashed, dotted, and solid lines in the photocenter and FWHM panels. A disk radius of 0.45 pixels, brightness contrast c$_{cont}$ = 10 magnitudes, and a seeing of 5 pixels is assumed in all cases.}
\label{Fig:models4}
\end{figure}

Finally, the most complex scenario is when the center of the sphere/disk is displaced with respect to the star as this is a combination of the previous disk/sphere cases and the binary case explained above. The additional parameters needed for this model are the separation between the center of the disk/sphere and the point star, as well as the angle between the slit $PA$ and the line that connects the center of the sphere/disk and the star. In order to show the influence of these parameters on the spectro-astrometric results, Fig. \ref{Fig:models4} shows the simple case when the slit is oriented towards the major and minor axes of the disk. If the center of the disk is shifted with respect to the star along the disk major axis, a photocenter shift appears when the slit is oriented in this direction, but no photocenter displacement is shown in the perpendicular one. On the contrary, if the disk center is displaced along the minor axis, a photocenter shift will only be apparent if the slit is oriented in this direction. When the disk is centered 45$\degr$ from the major and minor axes, a photocenter shift is apparent for both slit orientations associated with the corresponding projections. The FWHM signal is smaller when the slit is oriented along the minor axis and remains practically unchanged for all displacements of the disk centers as that  mainly reflects the disk extent at each slit orientation. Significantly, the appearance of a FWHM signal is always accompanied by a larger photocenter signature, which makes a displacement of disk center irrelevant for explaining our observations.  

\end{appendix}

\end{document}